\documentclass[11pt]{article}
\usepackage[russian]{babel}
\usepackage{amssymb,amsmath,amsthm}
\usepackage{graphicx}
\newcommand{\bc}{\begin{center}}
\newcommand{\ec}{\end{center}}
\newcommand{\be}{\begin{equation}}
\newcommand{\ee}{\end{equation}}
\newcommand{\bea}{\begin{eqnarray}}
\newcommand{\eea}{\end{eqnarray}}
\newcommand{\ba}{\begin{array}}
\newcommand{\ea}{\end{array}}
\newcommand{\edc}{\end{document}}

\def\l{\lambda}

\begin{document}
УДК 517.98
\begin{center}
\textbf{\Large {Единственность трансляционно-инвариантной меры
Гиббса для HC-моделей с четырьмя состояниями на
дереве Кэли}}\\
\end{center}

\begin{center}
Р.М.Хакимов\footnote{Институт математики, ул. Дурмон йули, 29, Ташкент, 100125, Узбекистан.\\
E-mail: rustam-7102@rambler.ru}
\end{center}

\begin{abstract} В этой работе рассмотрены плодородные Hard-Core (HC) модели
с параметром активности $\lambda>0$ и четырьмя состояниями на
дереве Кэли порядка два. Известно, что существуют три типа таких
моделей. В данной работе для каждой из этих моделей доказана
единственность трансляционно-инвариантной меры Гиббса.
\end{abstract}

\textbf{Ключевые слова}: дерево Кэли, конфигурация, НС-модель,
мера Гиббса, трансляционно-инвариантные меры.

\section{Введение}\

Изучение предельных мер Гиббса играет важную роль во многих
разнообразных областях науки. В частности, Hard-Core (жесткая
сердцевина) модель возникает при изучении случай-ных независимых
множеств графа (\cite{Gal}), при изучении молекул газа на решетке
(\cite{Ba}) и при изучении задач телекоммуникации (\cite{Kel1}).

Hard-Core(жесткий диск, твердое ядро, жесткая сердцевина) модель
на $D$-мерной решетке $ \mathbb Z ^ D$ была введена и изучена
Мазелью и Суховым в работе \cite{Maz}.

Описание всех предельных мер Гиббса, в частности,
трансляционно-инвариантных мер Гиббса, для данного гамильтониана
является одним из основных задач теории гиббсовских мер.
Определение меры Гиббса и понятия, связанные с этой теорией,
вводятся стандартным образом (см. например \cite {6}-\cite {Si}).

В работе \cite{7} изучена HC (Hard Core)-модель с двумя
состояниями на дереве Кэли и доказано, что
транцляционно-инвариантная мера Гиббса для этой модели
единственна. Кроме того, при некоторых условиях на параметры
НС-модели доказана неединственность периодических мер Гиббса с
периодом два. В работе \cite{XR} изучены слабо периодические меры
Гиббса для HC-модели с двумя состояниями для нормального делителя
индекса два и показана единственность
(трансляционно-инвариантность) слабо периодической меры для
HC-модели.

В работах \cite{MRS},\cite{Ro} были изучены гиббсовские меры для
(hard core) HC-модели с тремя состояниями на дереве Кэли порядка
$k\geq1$. В работе \cite{bw} выделены плодородные HC модели,
соответствующие графам "петля"\,, "свисток"\,, "жезл"\, и "ключ".
В работе \cite{MRS} были изучены трансляционно-инвариантные и
периодические меры Гиббса для HC модели в случае "ключ"\, на
дереве Кэли и была доказана единственность
трансляционно-инвариантной меры Гиббса для любой положительной
активности $\lambda$. В \cite{Ro} были изучены
трансляционно-инвариантные и периодические меры Гиббса для HC
модели в случаях "свисток"\,, "жезл"\, и "петля"\, и показано, что
(случай "петля") из положительной активности не следует
единственность трансляционно-инвариантных мер.

В настоящей работе рассматриваются плодородные HC-модели с
четырьмя состояниями, соответствующие графам "палка", "ключ" и
"пистолет", на однородном дереве Кэли порядка два. В каждом случае
доказана, что трансляционно-инвариантная HC-мера Гиббса
единственна.

\section{Определения и известные факты}\

Дерево Кэли $\Im^k$ порядка $ k\geq 1 $ - бесконечное дерево, т.е.
граф без циклов, из каждой вершины которого выходит ровно $k+1$
ребер. Пусть $\Im^k=(V,L,i)$, где $V$ есть множество вершин
$\Im^k$, $L$ - его множество ребер и $i$ - функция инцидентности,
сопоставляющая каждому ребру $l\in L$ его концевые точки $x, y \in
V$. Если $i (l) = \{ x, y \} $, то $x$ и $y$ называются  {\it
ближайшими соседями вершины} и обозначаются через $l = \langle x,
y \rangle $. Расстояние $d(x,y), x, y \in V$ на дереве Кэли
определяется формулой
$$
d (x, y) = \min \ \{d | \exists x=x_0, x_1,\dots, x _ {d-1},
x_d=y\in V \ \ \mbox {такой, что} \ \ \langle x_0,
x_1\rangle,\dots, \langle x _ {d-1}, x_d\rangle\} .$$

Для фиксированого $x^0\in V$ обозначим $ W_n = \ \{x\in V\ \ | \ \
d (x, x^0) =n \}, $
$$
 V_n = \ \{x\in V\ \ | \ \ d (x, x^0) \leq n \},\ \ L_n = \ \{l =
\langle x, y\rangle \in L \ \ | \ \ x, y \in V_n \}.
$$
Рассмотрим HC-модель ближайших соседей с четырмя состояниями на
однородном дереве Кэли. В этой модели каждой вершине $x$ ставится
в соответствие одно из значений $\sigma (x)\in \{0,1,2,3\}$.
Значения $\sigma (x)=1,2,3$ означают, что вершина $x$ `занята', а
значение $\sigma (x)=0$ означает, что вершина $x$ `вакантна'.

Конфигурация $\sigma=\{\sigma(x),\ x\in V\}$ на дереве Кэли
задается как функция из $V$ в $\{0,1,2, 3\}$. Множество всех
конфигураций на $V$ обозначается через $\Omega$. Аналогичном
образом можно определить конфигурации в $V_n$ ($W_n$), и множество
всех конфигураций в $V_n$ ($W_n$) обозначается как $\Omega_{V_n}$
($\Omega_{W_n}$).

Рассмотрим три типа плодородных (fertile) графов с четырмя
вершинами $0,1,2,3$ (на множестве значений $\sigma(x)$), которые
имеют следующие виды:

\[
\begin{array}{ll}
\mbox{\it палка(stick)}: &  \{0,1\}\{1,2\}\{2,3\};\\
\mbox{\it ключ(key)}: &  \{0,1\}\{0,2\}\{1,2\}\{2,3\};\\
\mbox{\it пистолет(gun)}: &  \{0,1\}\{0,2\}\{1,2\}\{2,2\}\{2,3\};\\
\end{array} \]

Графы, которые не являются плодородными, называются бесплодными
(sterile) (см.\cite{bw}).

Пусть $O=\{\textit{палка, ключ, пистолет}\}, \ G\in O.$
Конфигурация $\sigma$ называется $G$-\textit{допустимой
конфигурацией} на дереве Кэли (в $V_n$ или $W_n$), если $\{\sigma
(x),\sigma (y)\}$-ребро $G$ для любой ближайшей пары соседей $x,y$
из $V$ (из $V_n$). Обозначим множество $G$-допустимых конфигураций
через $\Omega^G$ ($\Omega_{V_n}^G$).

Множество активности \cite{bw} для графа $G$ есть функция $\l:G
\to R_+$ из множества $G$ во множество положительных
действительных чисел. Значение $\l_i$ функции $\l$ в вершине
$i\in\{0,1,2,3\}$ называется ее ``активностью''.

Будем писать $x<y,$ если путь от $x^0$ до $y$ проходит через $x$.
Вершина $y$ называется прямым потомком $x$, если $y>x$ и $x, \ y$
являются соседями. Через $S(x)$ обозначим множество прямых
потомков $x.$ Заметим, что в $\Im^k$ всякая вершина $x\neq x^0$
имеет $k$ прямых потомков, а вершина $x^0$ имеет $k+1$ потомков.

Для $\sigma_n\in\Omega_{V_n}^G$ положим
$$\#\sigma_n=\sum\limits_{x\in V_n}{\mathbf 1}(\sigma_n(x)\geq 1)$$
число занятых вершин в $\sigma_n$.

Пусть $z:\;x\mapsto z_x=(z_{0,x}, z_{1,x}, z_{2,x}, z_{3,x}) \in
R^4_+$ векторнозначная функция на $V$. Для $n=1,2,\ldots$ и $\l>0$
рассмотрим вероятностную меру $\mu^{(n)}$ на $\Omega_{V_n}^G$,
определяемую как
\begin{equation}\label{rus2.1}
\mu^{(n)}(\sigma_n)=\frac{1}{Z_n}\lambda^{\#\sigma_n} \prod_{x\in
W_n}z_{\sigma(x),x}.
\end{equation}

Здесь $Z_n$-нормирующий делитель:
$$
Z_n=\sum_{{\widetilde\sigma}_n\in\Omega^G_{V_n}}
\lambda^{\#{\widetilde\sigma}_n}\prod_{x\in W_n}
z_{{\widetilde\sigma}(x),x}.
$$
Говорят, что вероятностная мера $\mu^{(n)}$ является
согласованной, если $\forall$ $n\geq 1$ и
$\sigma_{n-1}\in\Omega^G_{V_{n-1}}$:
\begin{equation}\label{rus2.2}
\sum_{\omega_n\in\Omega_{W_n}}
\mu^{(n)}(\sigma_{n-1}\vee\omega_n){\mathbf 1}(
\sigma_{n-1}\vee\omega_n\in\Omega^G_{V_n})=
\mu^{(n-1)}(\sigma_{n-1}).
\end{equation}
В этом случае существует единственная мера $\mu$ на $(\Omega^G,
\textbf{B})$ такая, что для всех $n$ и $\sigma_n\in
\Omega^G_{V_n}$
$$\mu(\{\sigma|_{V_n}=\sigma_n\})=\mu^{(n)}(\sigma_n),$$
где $\textbf{B}-$$\sigma$-алгебра, порожденная цилиндрическими
подмножествами $\Omega^G$.

\textbf{Определение.} Мера $\mu$, определенная формулой
(\ref{rus2.1}) с условием (\ref{rus2.2}), называется
($G$-)HC-\textit{мерой Гиббса} с $\lambda>0$,
\textit{соответствующей функции} $z:\,x\in V
\setminus\{x^0\}\mapsto z_x$. Множество таких мер (для
всевозможных $z$) обозначается через ${\mathcal S}_G$.

Пусть $L(G)$-множество ребер графа $G$, обозначим через $A\equiv
A^G=\big(a_{ij}\big)_{i,j=0,1,2,3}$ матрицу смежности $G$, т.е.
$$ a_{ij}\equiv a^G_{ij}=\left\{\begin{array}{ll}
1,\ \ \mbox{если}\ \ \{i,j\}\in L(G),\\
0, \ \ \mbox{если} \ \  \{i,j\}\notin L(G).
\end{array}\right.$$

В следующей теореме сформулировано условие на $z_x$, гарантирующее
согласованность меры $\mu^{(n)}$.

\textbf{Теорема 1.}\cite{Rb} \label{rust1} Вероятностные меры
$\mu^{(n)}$, $n=1,2,\ldots$, заданные формулой (\ref{rus2.1}),
согласованны тогда и только тогда, когда для любого $x\in V$ имеют
место следующие равенства:
$$
z'_{0,x}=\lambda \prod_{y\in S(x)}{a_{10}+
a_{11}z'_{1,y}+a_{12}z'_{2,y}+a_{13}z'_{3,y}\over
a_{00}+a_{01}z'_{1,y}+a_{02}z'_{2,y}+a_{03}z'_{3,y}},$$
$$z'_{1,x}=\lambda \prod_{y\in S(x)}{a_{20}+
a_{21}z'_{1,y}+a_{22}z'_{2,y}+a_{23}z'_{3,y}\over
a_{00}+a_{01}z'_{1,y}+a_{02}z'_{2,y}+a_{03}z'_{3,y}},$$
$$z'_{2,x}=\lambda \prod_{y\in S(x)}{a_{30}+
a_{31}z'_{1,y}+a_{32}z'_{2,y}+a_{33}z'_{3,y}\over
a_{00}+a_{01}z'_{1,y}+a_{02}z'_{2,y}+a_{03}z'_{3,y}},$$ где
$z'_{i,x}=\lambda z_{i,x}/z_{3,x}, \ \ i=0,1,2$.\

\section{Трасляционно-инвариантные меры Гиббса}\

Мы полагаем, что $z_{3,x}\equiv 1$ и $z_{i,x}=z'_{i,x}>0,\ \
i=0,1,2$. Тогда для любых функций $x\in V\mapsto
z_x=(z_{0,x},z_{1,x},z_{2,x})$, удовлетворяющих равенству

\begin{equation}\label{rus3.1}
z_{i,x}=\lambda \prod_{y\in S(x)}{a_{i0}+
a_{i1}z_{1,y}+a_{i2}z_{2,y}+a_{i3}z_{3,y}\over
a_{00}+a_{01}z_{1,y}+a_{02}z_{2,y}+a_{03}z_{3,y}}, \ \ i=0,1,2,
\end{equation}
существует единственная $G$-HC-мера Гиббса $\mu$ и наоборот.
Начнем с трансляционно-инвариантных решений, в которых $z_x=z\in
R^3_+$, $x\neq x_0$. В случаях $G=\textit{палка}$,
$G=\textit{ключ}$ и $G=\textit{пистолет}$ из (\ref{rus3.1})
получим следующие системы уравнений:
\begin{equation}\label{rus3.01} \left\{\begin{array}{ll}
z_0=\lambda\left({ z_1\over z_2}\right)^k,\\[2mm]
z_1=\lambda\left({z_0+z_2\over z_2}\right)^k,\\[2mm]
z_2=\lambda\left({z_1+1\over z_2}\right)^k,
\end{array}\right.
\end{equation}

\begin{equation}\label{rus3.02} \left\{\begin{array}{ll}
z_0=\lambda\left({ z_1+z_2\over z_2}\right)^k,\\[2mm]
z_1=\lambda\left({z_0+z_2\over z_2}\right)^k,\\[2mm]
z_2=\lambda\left({z_0+z_1+1\over z_2}\right)^k,
\end{array}\right.
\end{equation}
\begin{equation}\label{rus3.03} \left\{\begin{array}{ll}
z_0=\lambda\left({ z_1+z_2\over z_2}\right)^k,\\[2mm]
z_1=\lambda\left({z_0+z_2\over z_2}\right)^k,\\[2mm]
z_2=\lambda\left({z_0+z_1+z_2+1\over z_2}\right)^k,
\end{array}\right.
\end{equation}
соответственно.

\textbf{Лемма.} \textit{В системах уравнений (\ref{rus3.02}) и
(\ref{rus3.03}) $z_0=z_1.$}

Доказательство непосредственно получается вычитанием из первого
уравнения второго в (\ref{rus3.02}) и (\ref{rus3.03}),
соответственно.\

Следующее утверждение дает оценки для произвольного решения
системы уравнений (\ref{rus3.01}):

\textbf{Утверждение 1.} Если $z=(z_0, z_1, z_2)$ является решением
(\ref{rus3.01}), то

$1. \ {\lambda^{k+1}\over
\sqrt[k+1]{\lambda^k(2^k\lambda+1)^{k^2}}}<z_0<{2^{k^2}\lambda^{k+1}\over
\sqrt[k+1]{\lambda^k(\lambda+1)^{k^2}}}; \ \ \ 2. \ \
\lambda<z_1<2^k\lambda;$

$3. \
\sqrt[k+1]{\lambda(\lambda+1)^k}<z_2<\sqrt[k+1]{\lambda(2^k\lambda+1)^k}$.

\textbf{Доказательство.} 1. Получается из первого уравнения
(\ref{rus3.01}), используя оценки для $z_1$ и $z_2$.

2. Из второго уравнения (\ref{rus3.01}) получим, что
$z_1>\lambda.$ Разделив первое уравнение на третье, будем иметь:
$${z_0\over z_2}=\left({z_1\over z_1+1}\right)^k<1.$$
Используя последнее неравенство, из второго уравнения следует, что
$z_1<2^k\lambda.$

3. Используя оценку для $z_1$, получается из третьего уравнения
(\ref{rus3.01}) $z_2=\sqrt[k+1]{\lambda(z_1+1)^k}$. Утверждение
доказано.

\subsection{Случай $G=\textit{палка}$}\

В этом случае верна следующая

\textbf{Теорема 2.} \textit{При $k=2$ и $\lambda
>0$ в случае $G=\textit{палка}$ существует
только одна Hard-Core трансляционно-инвариантная мера Гиббса.}\

\textbf{Доказательство.} Из третьего уравнения системы
(\ref{rus3.01}) при $k=2$ найдем $z_2$, и подставим в первое
уравнение. Полученные выражения для $z_2$ и $z_0$ подставим во
второе уравнение. В результате получим уравнение
\begin{equation}\label{rus3.5}
z_1=\lambda\cdot \left({z_1^2\over (z_1+1)^2}+1\right)^2=f(z_1).
\end{equation}
Так как
\begin{equation}\label{rus3.6}
f'(z_1)=4\lambda \left({z_1^2\over (z_1+1)^2}+1\right){z_1\over
(z_1+1)^3}>0,
\end{equation} то функция $f(z_1)$ строго возрастает, отсюда она не
имеет периодических точек
$$\lim_{n\rightarrow\infty}f^{(n)}(z_1)\neq z_1,$$
т.е. все предельные точки неподвижные.

Далее, из уравнения (\ref{rus3.5}) найдем $\lambda$ и подставим в
(\ref{rus3.6}). Тогда будем иметь
$$f'(z_1)={4z_1^2\over (z_1^2+(z_1+1)^2)(z_1+1)}.$$
Легко доказать, что $f'(z_1)<1$ для любых $z_1>0$. Из этого
следует, что все неподвижные точки притягивающие. Из всего
сказанного можно заключить, что уравнение $z_1=f(z_1)$ имеет
единственное решение при $\lambda>0$. Теорема доказана.

\subsection{Случай $G=\textit{ключ}$}\

В этом случае обозначая $\sqrt[k]{z_1}=x>0, \ \sqrt[k]{z_2}=y>0,
\sqrt[k]{\lambda}=a$ и используя лемму, из системы уравнений
(\ref{rus3.02}) получим следующую систему уравнений:

\begin{equation}\label{rus3.3} \left\{\begin{array}{ll}
x=a\cdot{x^k+y^k\over y^k}=a\cdot \left({x\over y}\right)^k+a,\\[2mm]
y=a\cdot{2x^k+1\over y^k}=2a\cdot \left({x\over y}\right)^k+{a\over y^k}.\\
\end{array}\right.
\end{equation}
Из первого уравнения (\ref{rus3.3}) найдем $\left({x\over
y}\right)^k={x-a\over a}$ и подставим на второе уравнение
$y=2x-2a+{a\over y^k}$ и получим
$$x={y\over 2}-{a\over 2y^k}+a.$$
Используя последнее выражение для $x$, из второго уравнения
(\ref{rus3.3}) будем иметь
$$y^{k+1}=a\cdot \left[2\left({y\over 2}-{a\over 2y^k}+a\right)^k+1\right].$$
Это уравнение при $k=2$ эквивалентно уравнению
$$f(y)=2y^7-ay^6-4a^2y^5-(4a^3+2a)y^4+2a^2y^3+4a^3y^2-a^3=0,$$
которое по известной теореме Декарта о количестве положительных
корней многочлена имеет не более трех положительных решения. Кроме
того, $f(0)=-a^3<0$ и $f(y)\rightarrow +\infty$ при $y\rightarrow
+\infty$, т.е. уравнение $f(y)=0$ имеет по крайней мере одно
положительное решение.

Итак, справедливо следующее

\textbf{Утверждение 2.} \textit{Система уравнений (\ref{rus3.3})
при $k=2$ имеет по крайней мере одно и не более трех решений.}\

Покажем, что (\ref{rus3.3}) имеет только одно решение при любых
значениях $a>0$. Легко показать, что уравнение $z_1=f(z_1)$ имеет
более одного решения тогда и только тогда, когда уравнение
$z_1f'(z_1)=f(z_1)$ имеет более одного решения. Воспользуемся этим
свойством дважды. Для этого из системы уравнений (\ref{rus3.02})
при $k=2$ можем получить уравнение
$$z_1=\lambda\cdot \left({z_1+\sqrt[3]{\lambda(2z_1+1)^2}\over \sqrt[3]{\lambda(2z_1+1)^2}}\right)^2=
\lambda\cdot
\left({z_1\over\sqrt[3]{\lambda(2z_1+1)^2}}+1\right)^2=f(z_1),$$
производная правой части которого равна
$$f'(z_1)=2\lambda \left({z_1\over\sqrt[3]{\lambda(2z_1+1)^2}}+1\right){2z_1+3\over \sqrt[3]{\lambda(2z_1+1)^5}}.$$
Тогда из уравнения $z_1f'(z_1)=f(z_1)$ получим
$$z_1={\sqrt[3]{\lambda(2z_1+1)^5}\over 2z_1+5}=\varphi(z_1).$$
Вычисляя производную
$$\varphi'(z_1)={4\lambda (2z_1+1)(2z_1+11)\over 3(2z_1+5)^2\sqrt[3]{\lambda^2(2z_1+1)}},$$
рассмотрим уравнение $z_1\varphi'(z_1)=\varphi(z_1)$. Оно
эквивалентно следующему квадратному уравнению
$$4z_1^2-8z_1+15=0,$$
которое не имеет вещественных решений. Значит, уравнение
$z_1=f(z_1)$ имеет не более одного решения. Отсюда, в силу
утверждении 2, справедлива

\textbf{Теорема 3.} \textit{При $k=2$ и $\lambda
>0$ в случае $G=\textit{ключ}$ Hard-Core трансляционно-инвариантная мера Гиббса единственна.}

\subsection{Случай $G=\textit{пистолет}$}\

Заметив, что $z_0=z_1$, из (\ref{rus3.03}) получим следующую
систему уравнений:

\begin{equation}\label{rus3.4} \left\{\begin{array}{ll}
x=a\cdot{x^k+y^k\over y^k}=a\cdot \left({x\over y}\right)^k+a,\\[2mm]
y=a\cdot{2x^k+y^k+1\over y^k}=2a\cdot \left({x\over y}\right)^k+{a\over y^k}+a,\\
\end{array}\right.
\end{equation} где $\sqrt[k]{z_1}=x>0, \
\sqrt[k]{z_2}=y>0, \sqrt[k]{\lambda}=a$.

Ясно, что $x>a.$  Аналогично предыдущему случаю, из (\ref{rus3.4})
можем иметь
$$y^{k+1}=2a\cdot \left({y\over 2}-{a\over 2y^k}+{a\over2}\right)^k+ay^k+a.$$
Преобразуем это уравнение при $k=2$:
$$f(y,a)=2y^7-3ay^6-2a^2y^5-(a^3+2)y^4+2a^2y^3+2a^3y^2-a^3=0.$$
Анализируя последнее уравнение подобно предыдущему случаю, получим
следующее

\textbf{Утверждение 3.} \textit{Система уравнений (\ref{rus3.4})
при $k=2$ имеет по крайней мере одно и не более трех решений.}\

Покажем, что (\ref{rus3.4}) при $k=2$ имеет только одно решение
при любых значениях $a>0.$ Для этого  в этом случае из
(\ref{rus3.4}) можно получить уравнение
\begin{equation}\label{rus3.7}
x=a\cdot\left[\left({x^3\over 2x^3-ax^2+x-a}\right)^2+1
\right]=f(x).
\end{equation} Производная функции $f(x)$ равна
\begin{equation}\label{rus3.8}
f'(x)={2ax^5\over (2x^3-ax^2+x-a)^3}(-ax^2+2x-3a).
\end{equation}
Преобразуя уравнение $xf'(x)=f(x)$, получим уравнение
$$2x^6(-ax^2+2x-3a)=(x^6+(2x^3-ax^2+x-a)^2)(2x^3-ax^2+x-a),$$
которое не имеет решений при $a>{1\over \sqrt{3}}$. Отсюда
уравнение (\ref{rus3.7}) в силу Утверждении 3 имеет только одно
решение при $a>{1\over \sqrt{3}}.$

Рассмотрим случай $0<a\leq{1\over \sqrt{3}}.$ Из (\ref{rus3.8})
следует, что функция $f(x)$ имеет критические точки
$x_1={1-\sqrt{1-3a^2}\over a}$ и $x_2={1+\sqrt{1-3a^2}\over a}$
($a<x_1<x_2$ при $0<a\leq{1\over \sqrt{3}}$), убывает при
$a<x<x_1$, $x>x_2$ и возрастает при $x_1<x<x_2$. Отсюда
$x_1=x_{min}$ и $x_2=x_{max}$. Заметим, что значение функции
$f(x)$ в точке $a$ лежит над биссектрисой $y=x$, т.к. $f(a)=2a>a$.
Кроме того, значения функции $f(x)$ в точках $x_{min}$ и $x_{max}$
лежат под биссектрисой $y=x$, т.к. $f(x_{min})<x_{min}$ и
$f(x_{max})<x_{max}$ при $0<a\leq{1\over \sqrt{3}}.$ (Рис.1 и
Рис.2)

Из всего сказанного следует

\textbf{Теорема 4.} \textit{При $k=2$ и $\lambda
>0$ в случае $G=\textit{пистолет}$ Hard-Core трансляционно-инвариантная мера Гиббса единственна.}

\begin{center}
\includegraphics[width=10cm]{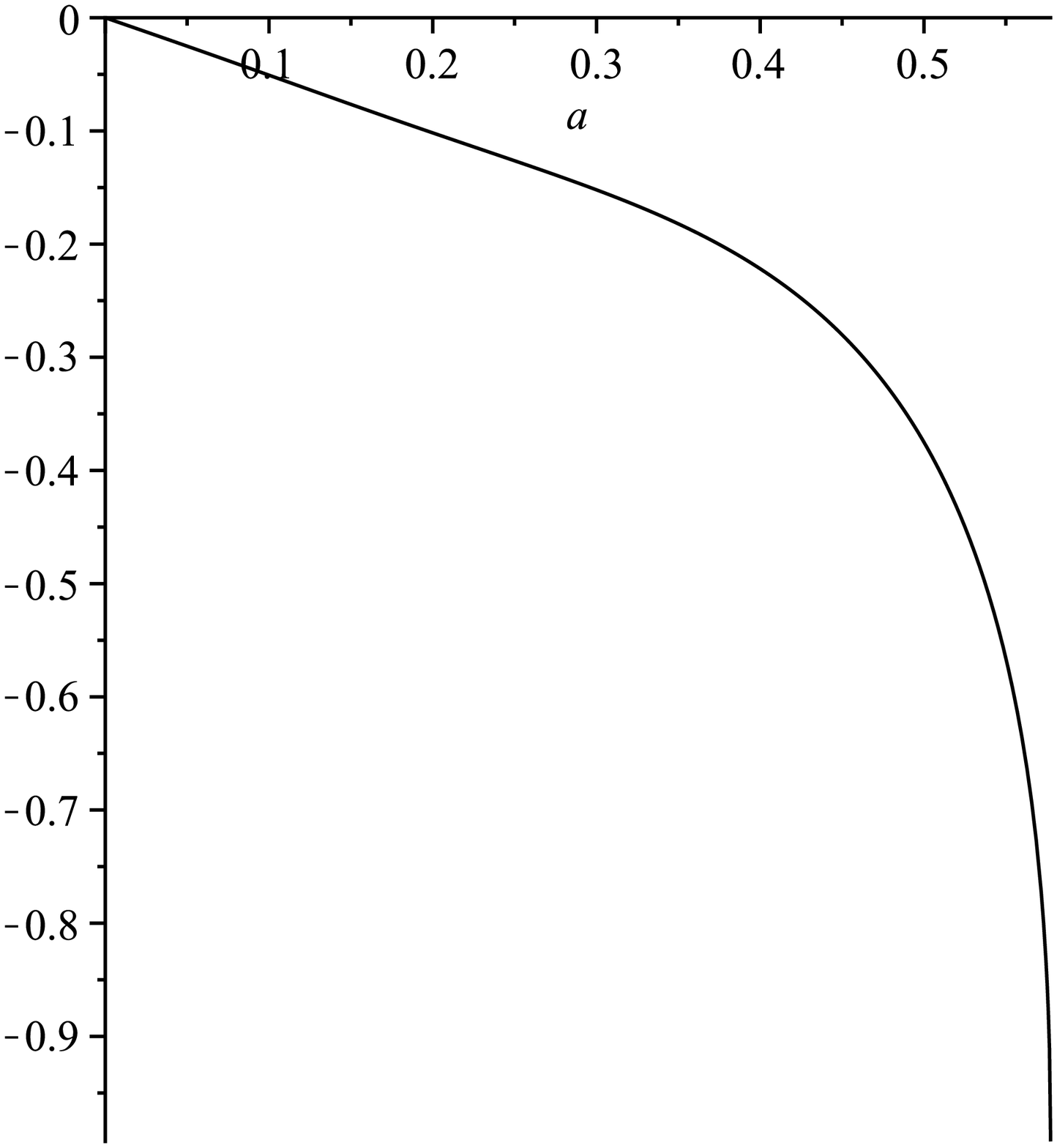}
\end{center}
\begin{center}{\footnotesize \noindent
 Рис.~1.
  График функции $g(a)=f(x_{min})-x_{min}$ при $0<a\leq{1\over \sqrt{3}}$}
\end{center}

\begin{center}
\includegraphics[width=10cm]{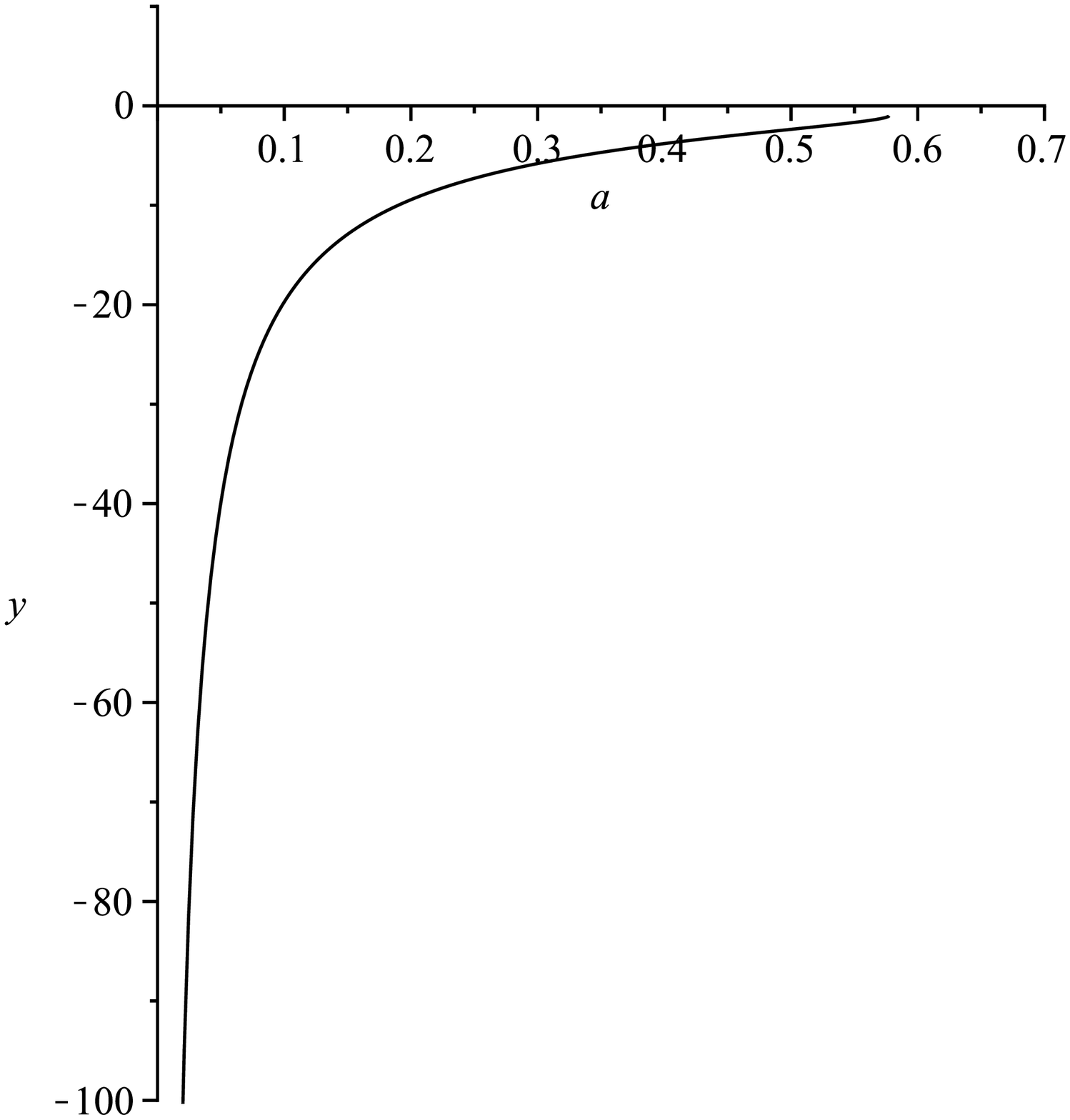}
\end{center}
\begin{center}{\footnotesize \noindent
 Рис.~2.
  График функции $g(a)=f(x_{max})-x_{max}$ при $0<a\leq{1\over \sqrt{3}}$}
\end{center}

\textbf{Благодарность.} Автор благодарит профессора У. А. Розикова
за постановку задачи и полезные советы.

\end{document}